\documentclass[preprint2,usenames,dvipsnames]{aastex6}

\usepackage[free-standing-units]{siunitx} 
\usepackage{natbib} 

\usepackage{multirow} 
\usepackage{paralist} 
\usepackage{xspace} 

\newcommand{\hh}{HAT-P-3b\xspace}
\newcommand{\tr}{TrES-3b\xspace}
\newcommand{\wf}{\textsc{Wide-FastCam}\xspace}
\newcommand{\seeonline}{See the online edition of the PASP for a color version
of this figure.}

\shorttitle{\hh and \tr transit observations}
\shortauthors{Ricci et al.}

\begin{document}

\title{Multi-filter transit observations of \hh and \tr\\
  with multiple Northern Hemisphere telescopes\footnote{Based upon
    observations acquired at: Observatorio Astron\'omico Nacional in
    the Sierra San Pedro M\'artir (OAN-SPM), Baja California,
    M\'exico; Observatorio de la Universidad de Monterrey (UDEM);
    Observatorio del Teide (OT); Osservatorio Regionale Parco
    Antola comune di Fascia (OARPAF).}}

\author{D.~Ricci\altaffilmark{1,2,3},
  P.~V.~Sada\altaffilmark{4},
  S.~Navarro-Meza\altaffilmark{3}, 
  R.~L\'opez-Valdivia\altaffilmark{5}, 
  R.~Michel\altaffilmark{3}, 
  L.~Fox~Machado\altaffilmark{3}, 
  F.~G.~Ram\'on-Fox\altaffilmark{6}, 
  C.~Ayala-Loera\altaffilmark{3,15}, 
  S.~Brown~Sevilla\altaffilmark{7},
  M.~Reyes-Ruiz\altaffilmark{3},
\and
  A.~La~Camera\altaffilmark{8},
  C.~Righi\altaffilmark{9,10},
  L.~Cabona\altaffilmark{9,10},
  S.~Tosi\altaffilmark{11,12}
\and
  N.~Truant\altaffilmark{13,1,2},
  S.~W.~Peterson\altaffilmark{14} 
  J.~Prieto-Arranz\altaffilmark{1,2},
  S.~Velasco\altaffilmark{1,2},
  E.~Pall\'e\altaffilmark{1,2},
  H.~Deeg\altaffilmark{1,2}
}

\altaffiltext{1}{
  Instituto de Astrof\'isica de Canarias,
  E-38205 La Laguna, Tenerife, Spain
\email{davide.ricci82@gmail.com}
}

\altaffiltext{2}{
  Universidad de La Laguna, Departamento de Astrof\'isica,
  E-38206 La Laguna, Tenerife, Spain
}

\altaffiltext{3}{
  Observatorio Astron\'omico Nacional, Instituto de
  Astronom\'ia -- Universidad Nacional Aut\'onoma de M\'exico,
  Ap. P. 877, Ensenada, BC 22860, M\'exico
}

\altaffiltext{4}{ Departamento de F\'isica y Matem\'aticas --
  Universidad de Monterrey, Avenida Ignacio Morones Prieto 4500 Pte.,
  Jes\'us M. Garza, 66238 San Pedro Garza Garc\'ia, N.L., M\'exico
}

\altaffiltext{5}{
  Instituto Nacional de Astrof\'isica, \'Optica y Electr\'onica, 
  Luis Enrique Erro 1, Tonantzintla, Puebla 72840 M\'exico
}

\altaffiltext{6}{
  SUPA, School of Physics and Astronomy, University
  of St Andrews, North Haugh, KY16 9SS, Scotland, UK
}

\altaffiltext{7}{ Facultad de Ciencias F\'isico-Matem\'aticas,
  Benem\'erita Universidad Aut\'onoma de Puebla,
  Av. San Claudio y 18 Sur, 72570 Puebla, M\'exico.
}

\altaffiltext{8}{
  Dipartimento di Informatica, Bioingegneria, Robotica
  e Ingegneria dei Sistemi (DIBRIS), Universit\`a di Genova, Via
  Dodecaneso 35, 16146 Genova, Italy
}

\altaffiltext{9}{
  Universit\`a degli Studi dell'Insubria, via Valleggio 11, 22100 Como, Italy
}

\altaffiltext{10}{
  INAF-Osservatorio Astronomico di Brera, via Bianchi 46, 23807 Merate (LC), Italy
}

\altaffiltext{11}{
  Dipartimento di Fisica -- Universit\`a degli studi di Genova,
  Via Dodecaneso 33, 16146 Genova, Italy
}

\altaffiltext{12}{
  Istituto Nazionale di Fisica Nucleare -- Sezione di Genova, 
  Via Dodecaneso 33, 16146 Genova, Italy 
}

\altaffiltext{13}{
  Dipartimento di Fisica -- Universit\`a degli Studi di Trieste, 
  Via Alfonso Valerio, 2, 34127 Trieste, Italy
}
 
\altaffiltext{14}{
Kitt Peak National Observatory, National Optical Astronomy Observatory, 950 N 
 Cherry Ave., Tucson, AZ, 85719, USA
}

\altaffiltext{15}{
  Observat\'orio Nacional MCTIC, Rua Geral Jos\'e Cristino 77,
  Rio de Janeiro, RJ, 20921-400, Brasil.
}

\begin{abstract}
  We present a photometric follow-up of transiting exoplanets \hh and
  \tr, observed by using several optical and near-infrared filters,
  with four small-class telescopes ($D=36$--$152\centi\meter$) in the
  Northern Hemisphere. Two of the facilities present their first
  scientific results.  New 10 \hh light curves and new 26 \tr light
  curves are reduced and combined by filter in order to improve the
  quality of the photometry.  Combined light curves fitting is carried
  out independently by using two different analysis packages, allowing
  the corroboration of the orbital and physical parameters in the
  literature.
  Results find no differences in the relative radius with the
  observing filter.
  In particular, we report for \hh a first estimation of the
  planet-to-star radius $ R_p/R_* = 0.1112^{+0.0025}_{-0.0026}$ in the
  $B$ band which is coherent with values found in the $VRIz'JH$
  filters.
  Concerning \tr, we derive a value for the orbital period of
  $P=1.3061862 \pm 0.0000001$ days which shows no linear variations
  over nine years of photometric observations.

\end{abstract}



\section{Introduction}
\label{info}

\begin{table*}
  \caption{Main characteristics of  
    observing facilities and their photometric instruments and altitude on the map on the right.
    \label{tel}}
  \includegraphics[width=0.176\textwidth]{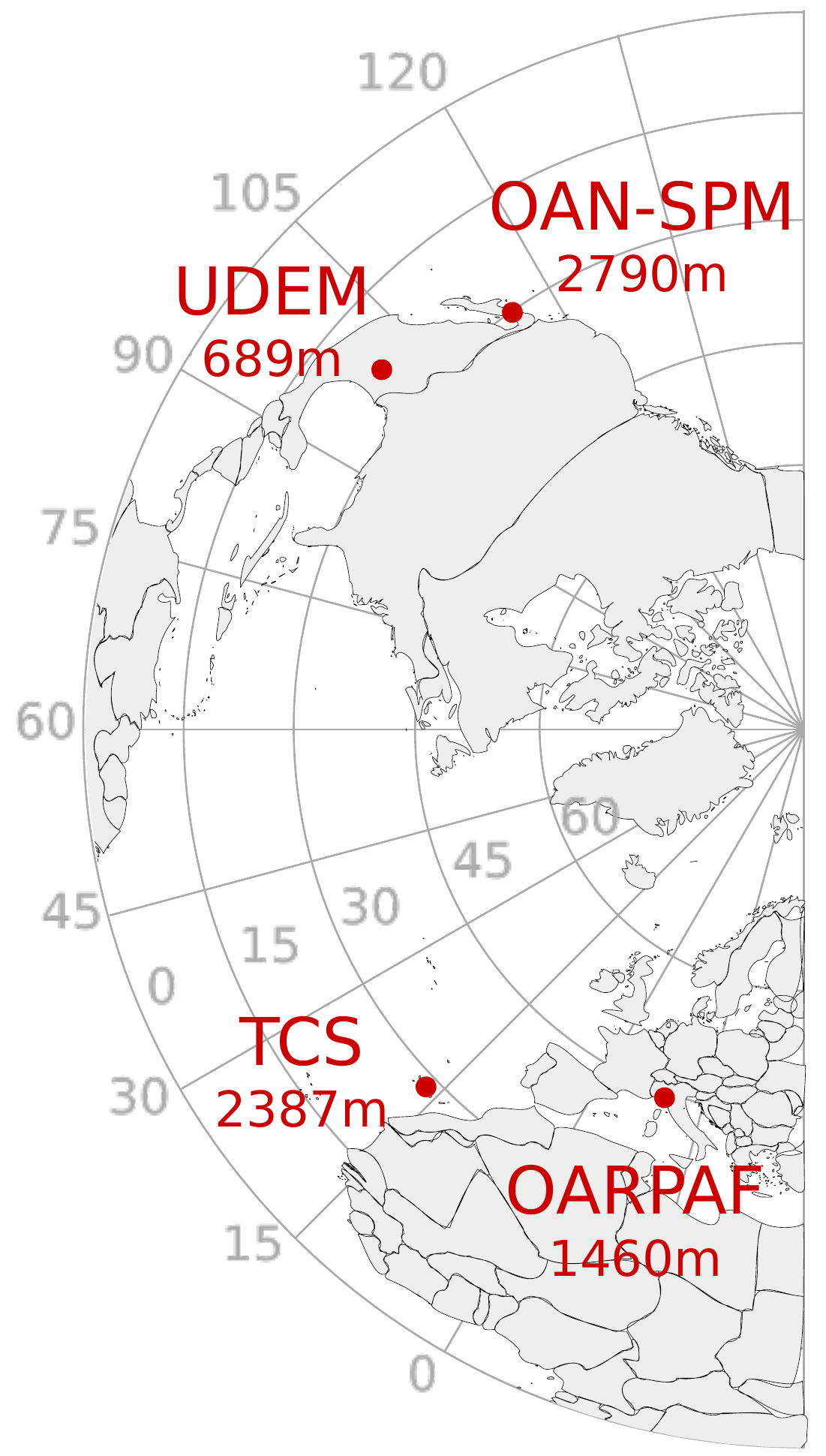}
\begin{minipage}[b]{0.80\textwidth}
  \flushright
  \begin{tabular}{lccccc}
\hline\hline
  Telescope    &              &     OAN-SPM      &          UDEM           &        IAC-TCS        & OARPAF           \\
\hline                                                                                        
   Diameter    & \centi\meter &        84        &           36            &          152          & 80             \\
 Focal ratio   &     $f/$     &       15.0       &          10.0           &         13.8          & 8              \\
   Latitude    &      N       &  \ang{31;02;39}  &    \ang{25;40;17.0}     &   \ang{28;18;01.8}    & \ang{44;35;27.9}      \\
  Longitude    &      W       & \ang{115;27;49}  &    \ang{100;18;31.0}    &   \ang{16;30;39.2}    & \ang{350;47;47.0}      \\
\hline                                                                                                 
  Instrument   &              & \textsc{Mexman}  & \textsc{Sbig stl-1301e} &          \wf          & \textsc{Sbig stl 11000m}  \\
\hline                                                                                                 
   CCD size    &      px      & 2043$\times$4612 &    1280$\times$1024     &   1024$\times$1024    & 2004$\times$1336      \\
  Resolution   &  arcsec/px   &       0.25       &           1.0           &         0.50          & 0.29            \\
Field of View  &    arcmin    & 8.4$\times$19.0  &    21.3$\times$17.1     &      8$\times$8       & 10$\times$10        \\
     Gain      &  $e^-$/ADU   &       1.8        &           2.3           &         0.10          & 0.8             \\
Read Out Noise &    $e^-$     &       4.8        &           16            & 2.7 (\texttt{g1m200}) & 12             \\
\hline
\end{tabular}
\end{minipage}
\end{table*}

The literature of exoplanetary transit observations started with the
works of \cite{2000ApJ...529L..45C} and \cite{2000ApJ...529L..41H} who
separately confirmed radial velocity shifts of the star \object{HD
  209458} \citep{1994SPIE.2198..362V, 1996AAS..119..373B} due to the
presence of its companion \object{HD 209458 b}.
Since then, big efforts have been made in order to build up
ground-based survey projects such as WASP \citep{2006PASP..118.1407P},
HAT \citep{2004PASP..116..266B} and its implementation HATSouth in the
Southern Hemisphere \citep{2011AAS...21725301P,2011AAS...21810304B},
TrES \citep{2007ASPC..366...13A}, XO \citep{2005PASP..117..783M}, KELT
\citep{2007PASP..119..923P} and the recent NGTS
\citep{2013EPJWC..4713002W}.

Among the current and forthcoming dedicated space-based programs we
mention Kepler \citep{2010Sci...327..977B}, PLATO \citep{plato}, and
TESS \citep{tess}, as well as the European CHEOPS mission
\citep{2014SPIE.9143E..2JF}, which will help to infer and characterize
the exoplanetary population, and has the ambition to detect exomoons
\citep{2015arXiv150800321S} and other interesting features.
Such a huge amount of data can benefit from useful online databases
such as ETD \citep{poddany10}\footnote{See for example observations
  provided by Phil Evans and flagged with a quality of ``2'' and ``3''
  \url{http://var2.astro.cz/ETD/etd.php?STARNAME=WASP-39\&PLANET=b},
  used to complement \cite{paper1-spm} data.}, and
\texttt{exoplanets.eu}\footnote{\url{http://exoplanets.eu}}
\citep{2011AA...532A..79S} to help astronomers in programming their
observations or defining best candidates for an independent follow-up.

In this framework, we have recently shown \citep{paper1-spm} the
adequacy of the \emph{San Pedro M\'artir -- Observatorio Astron\'omico
  Nacional} (OAN-SPM) facilities, located in Northwest Mexico, as a
valid resource in the Northern Hemisphere for the follow-up of
exoplanetary transits, particularly that of Hot Jupiters.  The survey,
which started in 2014 \citep{2014DPS....4621004R} and is still
ongoing, has allowed, to date, the observation of a total of 30
transiting exoplanetary
systems\footnote{\url{http://www.astrosen.unam.mx/~indy/spm-transits/}},
mainly with the $84\centi\meter$ telescope, achieving good quality
light curves in several filters, most of them through the Johnson's
$VRI$ filters.

In this paper we focus on two objects: the Jupiter-sized \hh and the
grazing transiting planet \tr, which we present in Sects.~\ref{hh}
and~\ref{tr}, respectively.
Moreover, we will present data from other sites by using not only
OAN-SPM survey data, but also light curves obtained from the following
facilities:
the \emph{Observatorio de la Universidad de Monterrey} (UDEM) in the
state of Nuevo L\'eon, Mexico;
the \emph{Telescopio Carlos S\'anchez} (TCS) at the \emph{Observatorio
  del Teide} (OT) in Tenerife, Spain, provided with \wf,
a new concept of fast camera with a wide Field of View (FoV);
and the \emph{Osservatorio Astronomico Regionale Parco Antola, comune
  di Fascia} (OARPAF) in northern Italy.
All the telescopes and the relative instruments involved in our
observations are described in Sect.~\ref{fac}.  In particular, \wf
is a new instrument, and OARPAF is a new facility, and we present
their first scientific results.
In Sects.~\ref{obs} and~\ref{data} we focus on observations and data
reduction of \hh and \tr.
Conclusions are described in Sect.~\ref{conc}.

\section{Targets}

\subsection{HAT-P-3}
\label{hh}

In the framework of the HATNet survey, \cite{2007ApJ...666L.121T}
discovered \object{\hh}, a Jupiter-size planet with a period of
$2.899703\pm0.000054$ days, which was the smallest of the 18
transiting extrasolar planets known at that time, with a planet radius
$R_p = 0.890\pm0.046 R_J$, where $R_J$ stands for the radius of
Jupiter.  \hh immediately appeared to be metal-rich, with a content
representing $75$ Earth masses $M_\earth$, (or, in Jupiter masses,
$0.24 M_J$).  Radial velocity measurements allowed authors to estimate
a planetary mass of $M_p = 0.599\pm0.026 M_J$, and found a distance
between planet and parent star of $0.03894\pm0.00070 \rm AU$, assuming
no orbital eccentricity.

System parameters with an improved precision were then updated by
\cite{2010MNRAS.401.1917G} by fitting seven light curves obtained
after performing aperture photometry on short-exposure
($4$--$5\second$), wide-band filter ($500$--$700\nano\meter$)
Liverpool Telescope data. Particular attention was given to possible
Earth-mass planets in the inner and outer $2:1$ orbital resonance, but
the lack of evidence of Transit Timing Variations (TTVs) excluded
this hypothesis.
Furthermore, \cite{2011AA...527A..85N, 2012MSAIS..19..105N} refined
the orbital parameters and ephemerides with a timing accuracy of
$11\second$, demonstrating the potential of the observing and
reduction strategy of the TASTE\footnote{The Asiago Search for Transit
  timing variations of Exoplanets} survey.

\begin{table}[t]
  \caption{Log of \hh observations. 
\label{loghh}}
\centering
\begin{tabular}{cCCc}
\tableline
\tableline
 Date (UT)  & \rm Filter & \rm Exp. time & Observatory         \\ 
\tableline
2009-01-19  &     Ic     &   60\second   & UDEM \\
2009-04-22  &     Ic     &   60\second   & UDEM \\
2009-05-15  &     JH     &   30\second   & KPNO\tablenotemark{a}\\
2009-05-15  &     z'     &   60\second   & KPNO-VCT\tablenotemark{a}\\
2009-05-21  &     Ic     &   60\second   & UDEM \\
2010-05-27  &     JH     &   20\second   & KPNO\tablenotemark{a} \\
2010-05-27  &     B      &   45\second   & KPNO-VCT\tablenotemark{a} \\
2012-04-22  &     Ic     &   60\second   & UDEM \\
2012-05-24  &     Ic     &   60\second   & UDEM \\
2013-06-02  &     Ic     &   60\second   & UDEM \\
2014-04-01  &     R      &  120\second   & OAN-SPM \\
2014-03-13  &     R      &  120\second   & OAN-SPM \\
2014-03-16  &     I      &  120\second   & OAN-SPM \\
2014-05-13  &     V      &   90\second   & OAN-SPM \\
\tableline
\end{tabular}
 \tablenotetext{a}{Published in
   \cite{2012PASP..124..212S}.} 
\end{table}
\begin{table}[t]
  \caption{Log of \tr observations. 
\label{logtr}}
\centering
\begin{tabular}{cCCc}
\tableline
\tableline
 Date (UT)  & \rm Filter & \rm Exp. time & Observatory         \\ 
\tableline                          
2007-07-23  &     Ic     &  120\second   & UDEM    \\
2007-08-09  &     Ic     &  120\second   & UDEM    \\
2008-07-11  &     Ic     &   60\second   & UDEM    \\
2008-07-15  &     V      &   90\second   & UDEM    \\
2008-10-04  &     V      &   90\second   & UDEM    \\
2009-05-06  &     z'     &   45\second   & KPNO-VCT\\
2009-05-10  &     z'     &  120\second   & KPNO-VCT\\
2009-05-31  &     Ic     &  120\second   & UDEM    \\
2009-07-17  &     z'     &  120\second   & UDEM    \\
2010-04-04  &     z'     &   40\second   & StPr    \\
2010-04-25  &     V      &   90\second   & UDEM    \\
2010-06-19  &     V      &   90\second   & UDEM    \\
2010-07-19  &     V      &   90\second   & UDEM    \\
2010-08-05  &     Ic     &   90\second   & UDEM    \\
2011-04-27  &     Ic     &   90\second   & UDEM    \\
2011-06-08  &     Rc     &   60\second   & UDEM    \\
2011-08-24  &     Rc     &   60\second   & UDEM    \\
2012-08-12  &     Rc     &   60\second   & UDEM    \\
2013-08-14  &     Rc     &   60\second   & UDEM    \\
2013-10-08  &     V      &   90\second   & UDEM    \\
2014-06-13  &     Rc     &   60\second   & UDEM    \\
2015-07-17  &     R      &  120\second   & TCS\tablenotemark{a}     \\
2015-08-03  &     R      &  120\second   & OARPAF  \\
2015-08-22  &     R      &  120\second   & OAN-SPM \\
2016-04-09  &     R      &  120\second   & OAN-SPM \\
2016-04-13  &     R      &  120\second   & OAN-SPM \\
2016-05-30  &     I      &   60\second   & OAN-SPM\tablenotemark{b} \\
2016-06-03  &     I      &   60\second   & OAN-SPM \\
\tableline
\end{tabular}
\tablenotetext{a}{binning 120 individual frames of $1\second$ each.}
\tablenotetext{b}{binning   6 individual frames of $10\second$ each.}
\end{table}

The anomalous radius of \hh, smaller than a pure hydrogen-helium
planet suggested the presence of a heavy-element core estimated to be
$100M_\earth$. This was the subject of the investigation of
\cite{2011AJ....141..179C, 2012AJ....144...90C}, who observed the
target in the $i$ and $z'$ bands, again finding no deviations in the
timing of eclipses. \cite{2011AJ....141..179C} found a period of
$P=2.8997360\pm0.0000020$ days, a planet-to-star radius ratio of
$R_p /R_*=0.1063\pm0.0020$, an orbital inclination of
$i=87.07\pm0.55\degree$, and a scaled semi-major axis of
$a/R_*=10.39\pm0.49$, adopting a circular orbit.
An additional follow-up at Kitt Peak National Observatory was carried
out by \cite{2012PASP..124..212S}, while \cite{2012MNRAS.426.1291S}
conducted a homogeneous study of more than 30 extrasolar planets
including \hh, based on all available data sets in the literature.
The author found a larger stellar radius ($0.947_{-0.027}^{+0.015}
M_J$) and pointed out that further photometric monitoring was needed
to get rid of the discrepancy between the data sets of
\cite{2007ApJ...666L.121T, 2010MNRAS.401.1917G, 2011AA...527A..85N}
and \cite{2011AJ....141..179C}.
Finally, a space-based investigation of secondary eclipses in
$3.6$--$4.5\micro\meter$ spectral bands of the \emph{Spitzer Space
  Telescope} was carried out by \cite{2013ApJ...770..102T}, finding
inefficiencies in heat redistribution, but letting the scenario open
about thermal inversion in the atmosphere. Moreover,
\cite{2013ApJ...770..102T} found the eccentricity of \hh consistent
with zero.

\subsection{TrES-3b}
\label{tr}

\object{\tr} is a massive transiting Hot Jupiter discovered by
\cite{2007ApJ...663L..37O} and confirmed by radial velocity
measurements.  Authors derived a period of $P=1.30619\pm0.00001$ days,
a semi-major axis of $a=0.0226\pm0.0013\rm AU$ and a near-grazing
inclination value of $i=82.15\pm0.21\degree$.  The
mass of the planet is estimated to be $M_p=1.92\pm0.23M_J$ in front of
a stellar mass of $M=0.9\pm0.15M_\sun$, while the planet radius is
$R_p=1.295\pm0.081R_J$.
\cite{2009AA...493L..35D} reported a temperature of the planet of
$T=2040\pm185\kelvin$ by measuring the thermal emission in the $K$
band using secondary eclipse observations which were carried out with
the William Hershel Telescope, that complemented the space-based
observations of the Spitzer Telescope.

The same value was found by \cite{2011ASPC..450...59D}.
\cite{2009AA...493L..35D} also suggest that the planet is in a
non-circular orbit.  Measurements of the radius of the planet in the
$K$ band do not significantly differ from the results obtained with
optical observations.

Many follow-ups were carried out by several groups in optical and UV
bands \citep{2011AAS...21812811T, 2012AAS...21933901J,
  2012AAS...21933909W, 2012IAUS..282..135V, 2012AAS...22012904S}. In
particular, \cite{2013MNRAS.428..678T} did not detect any early
ingress in UV as predicted by \cite{2011MNRAS.414.1573V}, resulting in
an abnormally small strength of the magnetic field.
The observations in the near-infrared by \cite{2009AA...493L..35D},
and the near UV by \cite{2013MNRAS.428..678T}, do not show differences
in the value of the radius of the planet.

An attempt to calculate TTVs was made for the first time by
\cite{2013ASInC...9...78T}, using 32 transits in the existing
literature, and 5 new transits. Authors use dynamic
models to suggest the presence of an additional outer planet close to
the $1:2$ resonance, with an estimated mass, in terms of Earth
masses, $\approx100M_\earth$.
The result contradicts that found by \cite{2013ApJ...764....8K}, which
observed an additional set of 11 transits in the framework of the
Apache Point Survey of Transit Lightcurves of Exoplanets (APOSTLE),
excluding the possibility of other planets and confirming the system's
parameters but with reduced error bars.
Finally, transit times of \tr were updated by
\cite{2013IBVS.6082....1M}, whose results also support no TTVs.

\section{Observations}

\subsection{Facilities}
\label{fac}

Observations presented in this paper were carried out by using four
different telescopes located in the Northern Hemisphere.  A summary of
their characteristics is shown in Table~\ref{tel}.

\begin{itemize}
  
\item 
  The first telescope is the OAN-SPM $84\centi\meter$.
  It provides a set of instruments that can be mounted according to
  the observational needs, among which \textsc{Mexman}, a wide-field
  imager already successfully tested for the observation and
  characterization of exoplanets using the transit method
  \citep{paper1-spm}.
  
\item The second telescope is the UDEM $36\centi\meter$, a
  \textsc{lx200gps} college telescope located close to the city of
  Monterrey, Mexico, and classed with a Minor Planet Center Code of
  720. It is available for student training and has been active for
  almost 10 years in the field of exoplanetary transit observations
  \cite{fox, 2016PASP..128b4402S}.
    
\item
  The third telescope, the TCS, is a $152\centi\meter$ equatorial
  Cassegrain
  telescope\footnote{\url{http://www.iac.es/telescopes/pages/es/inicio/telescopios/tcs.php}}
  located at the OT and managed by the Astrophysics Institute of
  Canary Islands (IAC), Spain.
  This telescope is provided with different instruments.  One of them,
  \textsc{FastCam}, allows quasi-diffraction limited real-time
  observations using the Lucky Imaging technique, the implementation
  of the instrument being targeted to high temporal resolution up to
  tens of images per second \citep{2008SPIE.7014E..47O}.
  Despite these advantages, \textsc{FastCam} is limited
  \citep{2010SPIE.7735E..3EM} by its $20\arcsec\times 20\arcsec$ FoV.
  
  For this reason, an implementation of the \textsc{FastCam} detector
  with a wider FoV was carried out: \wf offers a $1024\times1024$px
  EMCCD array and an optical design \citep{2014SPIE.9147E..6QM} to
  provide observers with an $\approx8\times8\arcmin$ FoV.
  We used \wf for our observations, in order to test its small readout
  time and low electronic noise.
  Linearity tests on the camera on flat field images show that the
  instrument works in the linear regime between 1700 and 4000 counts
  \citep{2016SPIE.9908E..2OV, 2017hsa9.conf..707V}. During
  observations we took care to tune the exposure time accordingly.   

  
\item The fourth telescope is the
  alt-azimuthal\footnote{\url{https://www.difi.unige.it/it/ricerca/altri-progetti/osservatorio-monte-antola}}
  OARPAF $80\centi\meter$, located near Mt.~Antola in Northern Italy,
  and whose scientific activity is managed by the Physics Department
  (DIFI) of the University of Genova, Italy.
  The telescope was designed by the Astelco
  company\footnote{\url{http://www.astelco.com}} to foresee a double
  focal station: the first, provided with a field derotator, is
  dedicated to scientific observations, and currently equipped with an
  air-cooled \textsc{Sbig stl 11000m} camera and a standard $UBVRI$
  Johnson filter wheel \citep{2012ASInC...7....7F}; the second is
  dedicated to ocular observations by amateurs.
  
  Pointing and positioning of the secondary mirror are controlled
  using the proprietary \texttt{AstelOS} software, provided by the
  constructor, on a dedicated Linux machine, while the image data
  capture is managed by the \texttt{MaxIm} software.
  The time stamp is obtained via a Global Positioning System
  (GPS) device.
  The instrument was recently fully-characterized with standard stars
  observations in all available filters for zero-point determination
  and extinction coefficient determination. Concerning the CCD
  commissioning, deep tests allowed calculation of: gain, Read Out
  Noise, dark current, plate scale, and linearity regime
  \citep{2016NCimC..39..284R}.
  
\end{itemize}

\subsection{Targets}
\label{obs}

We carried out 4 observations of \hh at OAN-SPM in nearly-full moon
conditions, in 2014, over a period spanning five months using three
filters: two with the $R$ filter, one with the $I$ filter and the
final one with the $V$ filter.  Except for the first $R$ observation,
we were able to follow the target for roughly five hours.
We also obtained 6 transits from UDEM taken with the $Ic$ filter
between 2009 and 2013.
Moreover, we also aggregated 4 already published observations from
\cite{2012PASP..124..212S}: two light curves from the Kitt Peak
National Observatory (KPNO) $200\centi\meter$ telescope observed with
the $JH$ filter, one light curve from the KPNO Visitor Center
Telescope (KPNO-VCT) $51\centi\meter$ telescope with the $B$ filter
and one curve from the same telescope with the Sloan $z'$ filter.
A log of \hh observations is shown in Table~\ref{loghh}.

Concerning \tr, we obtained three complete light curves in the $R$
band at OAN-SPM, and two in the $I$ band. Except for the first $I$
light curve, we used the defocused photometry method described in
detail by \cite{2014MNRAS.444..776S} and in previous papers of the
series.
We also present 18 UDEM transits: six with the $V$ filter, five with the
$Rc$ filter, six in the $Ic$ band and finally one in the $z'$
band. Additional $z'$ data come from KPNO (2 curves). An additional
curve in the $z'$ band comes from the KPNO-VCT operator Steven
Peterson (StPr) using his private observatory, and located close to
the KPNO facilities.
TCS also observed the target one night by using \wf and short-exposure
images of $1\second$ each, in the $R$ band.  Finally, one observation
with the $R$ filter was carried out at OARPAF.
We carefully checked that the targets and reference stars were not
relaying too close to hot or bad pixels or bad lines of devices, nor
at the edge of their FoV. We also used a $2\times2$ binning for our
observations in OAN-SPM, UDEM and OARPAF, windowing the devices in
order to minimize the readout time.
OAN-SPM, UDEM and OARPAF telescopes were also slightly defocused in
order to spread the Point Spread Function over a large number of
pixels, allowing longer exposure times, a shorter total readout time,
and reducing errors due to pixel response or flat-field correction
problems.
A log of \tr observations is shown in Table~\ref{logtr}.

\section{Data analysis}
\label{data}

Images of both targets were debiased and flat-field corrected.
We used the aperture photometry technique to obtain light curves.
In particular, we used the \texttt{defot} routine
\citep{2010MNRAS.408.1680S} written in the IDL language, that we
modified to work with FITS headers generated by the OAN-SPM, TCS
and OARPAF telescopes,
while UDEM data were reduced using independent custom IDL routines.
OAN-SPM and OARPAF images were not aligned, as we decided to calculate
and follow, for each image, the centroid of the target and that of the
reference stars, by using a cross-correlation method that we
re-implemented in \texttt{defot}.
For the OARPAF data coming from an Alt-Az telescope, we set up a
rotation tracking subroutine in order to correct for the residual
rotation of the field due to eventual errors in the alignment between
the optical axis, the derotator, and the CCD center.

Several field stars were tested in order to find a reference from
which performing differential photometry, and for both \hh and \tr we
chose a number between 2 and 8 non-saturated reference stars with
count values as close as possible to those of the target, also taking
into account sky conditions and availability.

A first-order trend in light curves was removed with the technique
described by \cite{fox}, in order to get rid of airmass effects.
Timestamps of the light curve points were converted to the
Dynamical Time-based system ($BJD_{\rm TDB}$) applying the transformation
given by \cite{eastman10}.

Data were then combined by using the procedure described by
\cite{2016PASP..128b4402S}.
Using this procedure, we obtained a total of 7 combined curves for \hh
for each one of the following filters:
$B$ (1 curve from KPNO-VCT), 
$V$ (1 curve from OAN-SPM),   
$R$ (2 curves from OAN-SPM),  
$I$ (1 curve from OAN-SPM), 
$Ic$ (6 curves from UDEM), 
$z'$ (1 curve from KPNO-VCT), 
$JH$ (2  curve from KPNO).

The same technique was applied to \tr data, and we obtained a total of
6 combined curves in the different filters as follows:
$V$ (6 curves from UDEM),
$R$ (3 curves from OAN-SPM, 1 from TCS, 1 from OARPAF),
$Rc$ (5 curves from UDEM),
$I$ (2 curves from OAN-SPM),
$Ic$ (6 curves from UDEM),
$z'$ (2 curves from KPNO, 1 from StPr, 1 from UDEM).


\begin{figure}[t]
  \centering
  \includegraphics[width=\columnwidth]{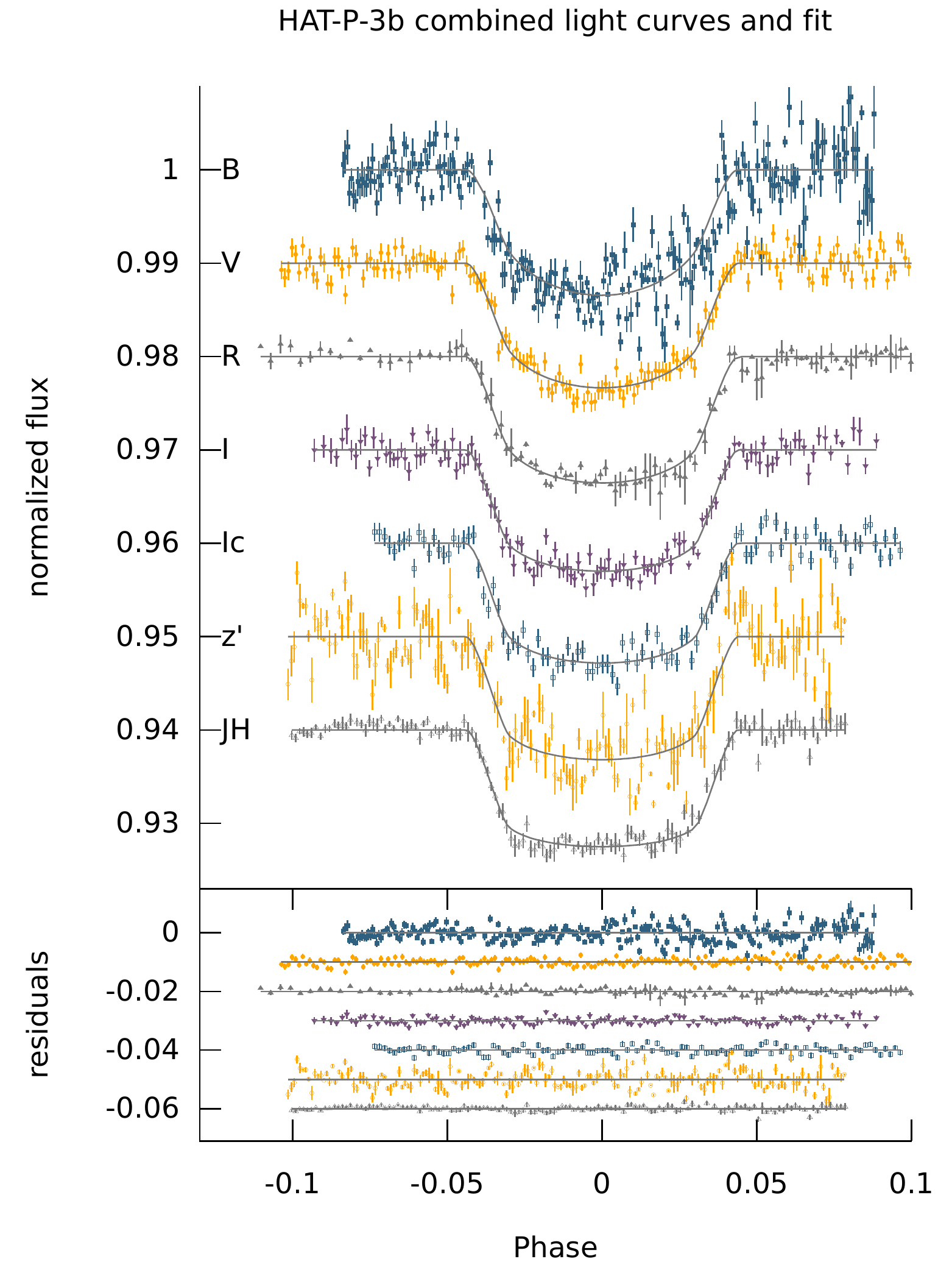}
  \caption{Combined light curves of \hh and fit models.
    Models are slightly shifted in flux for better
    visualization. The lower panel shows residuals.
    \seeonline
    \label{hat-sup}}
\end{figure}

Light curves were fitted with \textsc{exofast} \citep{eastman12,
  2013PASP..125...83E}.  A value of the inclination $i$, of the
semi-major axis in terms of the host star radius $a/R_*$, of the
Mid-time $T_{mid}$, and of the radius of the planet in terms of the
host star radius $R_p/R_*$ were fitted for each combined curve.
An independent light curve fitting was performed by using the IDL
software \emph{Transit Analysis Package} (\textsc{tap}) implemented by
\cite{gazak11}.
Both software use a Markov Chain Monte Carlo (MCMC) method to
find best fit parameters for the \cite{mandel02} model.

Differing from \textsc{exofast}, \textsc{tap} allows fitting
together, or separately, a set of parameters linking their value with
a lock matrix.  Thus it was possible to simultaneously fit a unique
value of the inclination $i$ and of the scaled semi-major axis $a/R_*$
for all combined curves, a different Mid-time $T_{mid}$ value for each
light curve, and a scaled planet radius $R_p/R_*$ value for each of
the different observing filter.

\subsection{\hh}
\label{hhfit}

In fitting procedures we fixed the period: $P =
2.8997382$ days \citep{2012PASP..124..212S}, and we supposed a circular
orbit (eccentricity $e=0$, argument of periastron $\omega=0$).
Concerning limb darkening, for \textsc{exofast} fitting we used
interpolated values from \cite{2000AA...363.1081C}, with the following
parameters:
$T_* = 5224\pm69\kelvin$, 
$\log{g_*} = 4.58\pm0.03$, and
 $[Fe/H]_* = 0.41 \pm 0.08$
 \citep{2007ApJ...666L.121T, 2012ApJ...757..161T}; while for
 \textsc{tap} fitting we used an online tool
 \citep{eastman12,eastman13} which interpolates atmosphere models of
 \cite{claret11}.

\begin{figure}[t]
  \centering
  \includegraphics[width=\columnwidth]{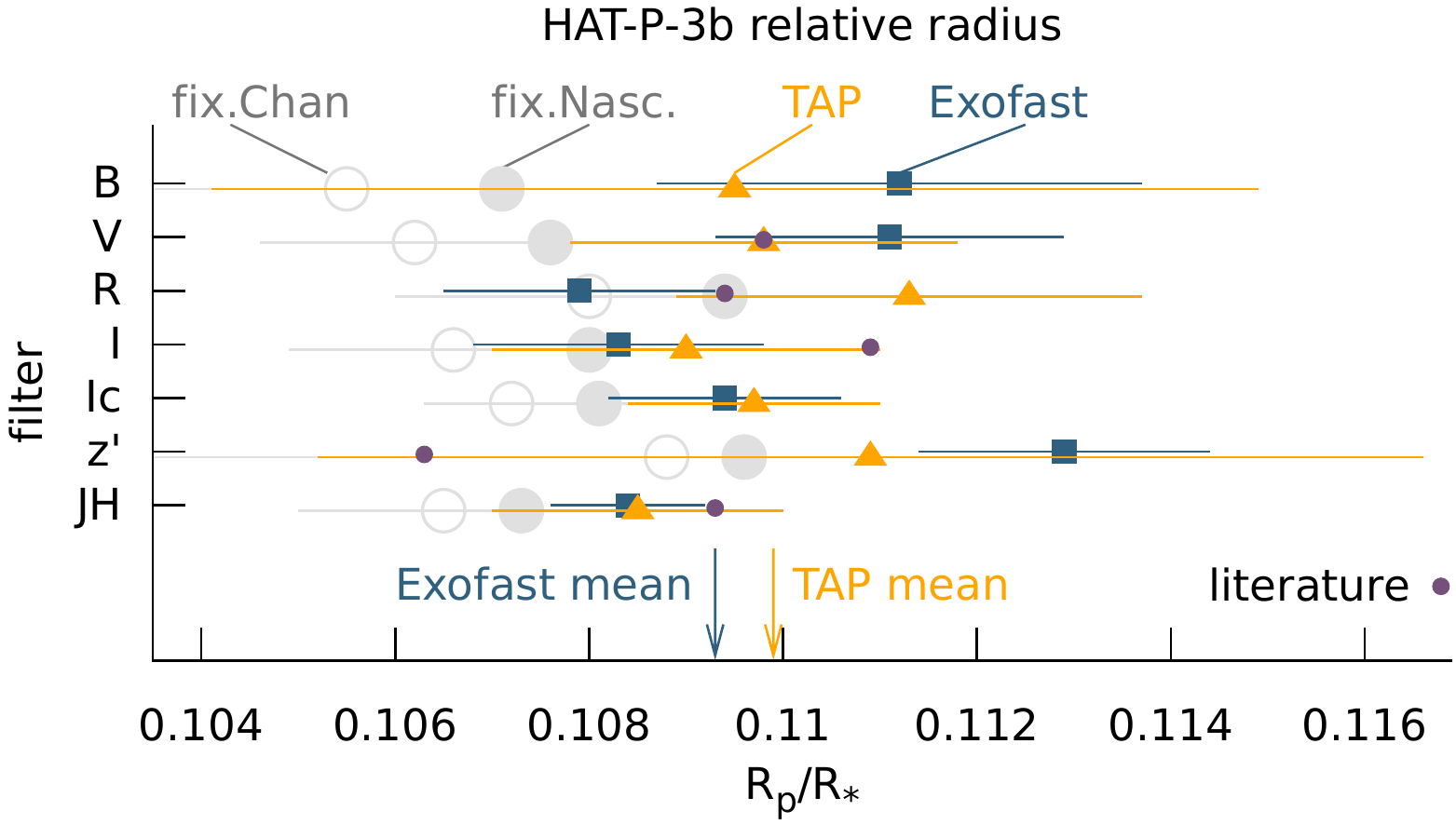}
  \caption{ $R_p/R_*$ values of \hh for each observing filter
    resulting from \textsc{exofast} and \textsc{tap} fit procedures.
    Arrows show weighted means.  
    Large light circles show \textsc{tap} results obtained by fixing
    $i$ and $a/R_*$ with values provided by \cite{2011AJ....141..179C}
    and \cite{2011AA...527A..85N}. Small bold
    circles show values provided in the literature:
    %
    $V+R$ \citep{2010MNRAS.401.1917G},
    $R$   \citep{2011AA...527A..85N},  
    $I$   \citep{2007ApJ...666L.121T}, 
    $z'$  \citep{2011AJ....141..179C},  and
    $JH$  \citep{2012PASP..124..212S}
    \seeonline
    \label{hat-radius}}
\end{figure}
\begin{figure}[b]
  \centering
  \includegraphics[width=\columnwidth]{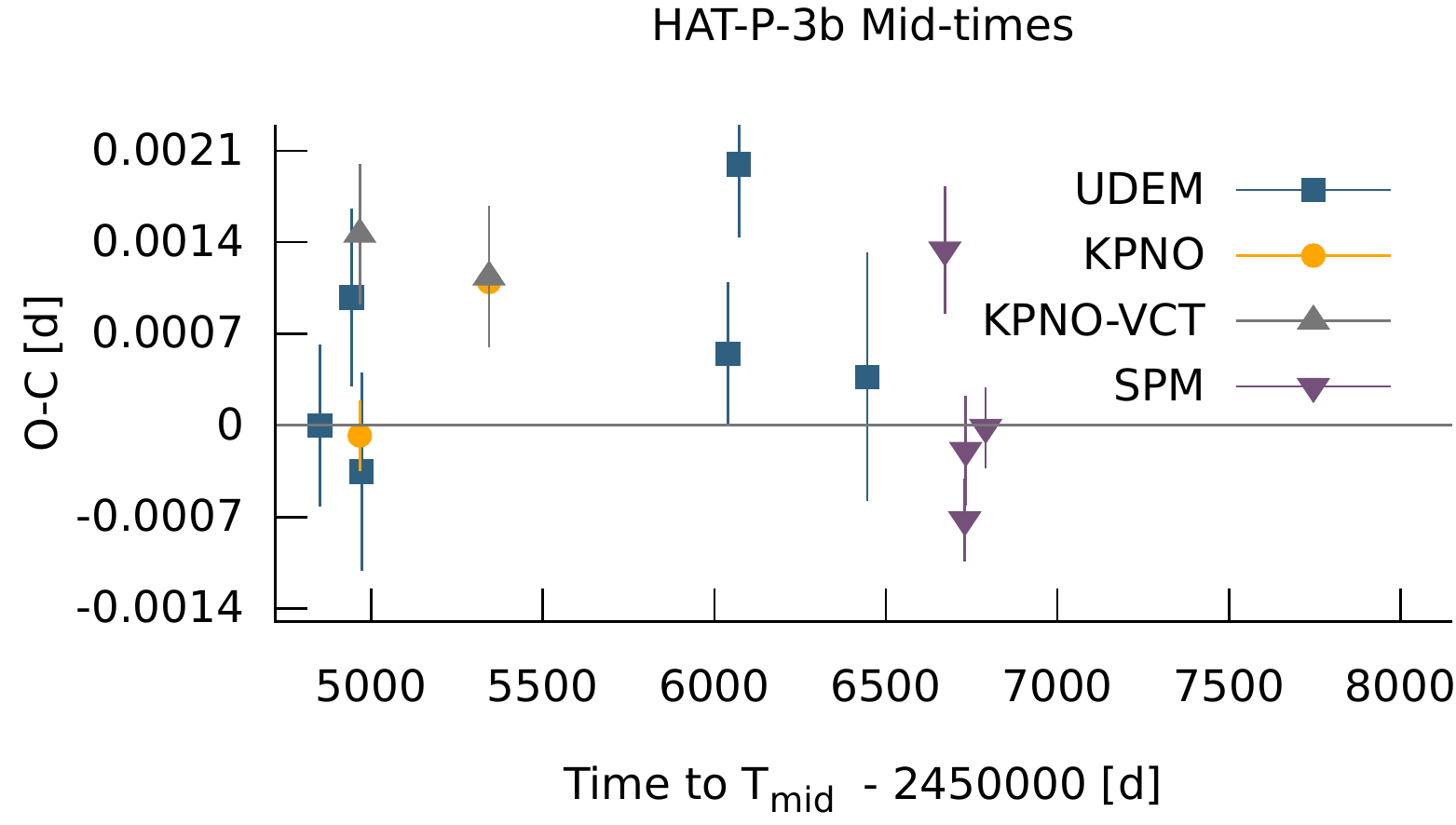}
  \caption{Observed minus Calculated Mid-times values for \hh
    observations presented in this work.
    \seeonline
    \label{hat-midtimes}}
\end{figure}

Fit results obtained with \textsc{exofast} are shown in
Table~\ref{fithh}.
In particular, weighted means obtained with \textsc{exofast}
are consistent with those of \cite{2011AA...527A..85N}:
\textsc{exofast} obtain
  $i       = 86.73^{+0.11}_{-0.11}$      against $86.75^{+0.10}_{-0.10}$ of \cite{2011AA...527A..85N};
$a/R_*     = 10.13^{+0.11}_{-0.11}$      against $10.12^{+0.32}_{-0.32}$; and
$R_p/R_*   = 0.1093^{+0.0005}_{-0.0005}$ against $0.1094^{+0.0011}_{-0.0011}$.
\textsc{tap} calculates a unique value for the first two parameters:
the inclination
$i= 86.27_{-0.18}^{+0.28}$ and the scaled semi-major axis
$a/R_*=9.65_{-0.18}^{+0.28}$,
which are slightly different from \textsc{exofast} and
\cite{2011AA...527A..85N} results, but in agreement within $3\sigma$.
A visual compairson between \textsc{exofast} and \textsc{tap} fit
values of the scaled planet radius $R_p/R_*$ is shown in
Fig.~\ref{hat-radius}.

%
Fig.~\ref{hat-radius} also show additional \textsc{tap} fit solutions
obtained by fixing $i$ and $a/R_*$ found by \cite{2011AJ....141..179C}
and \cite{2011AA...527A..85N}, and fitting only a separate value of
$R_p/R_*$ for each observing filter.
%
%
%
The figure gives evidence of the fact that
we find no significant radius variation with the observing filter.
In Fig.~\ref{hat-midtimes} calculated Mid-times are shown.

\clearpage

\subsection{\tr}
\label{trfit}

\begin{figure}[t]
  \centering
  \includegraphics[width=\columnwidth]{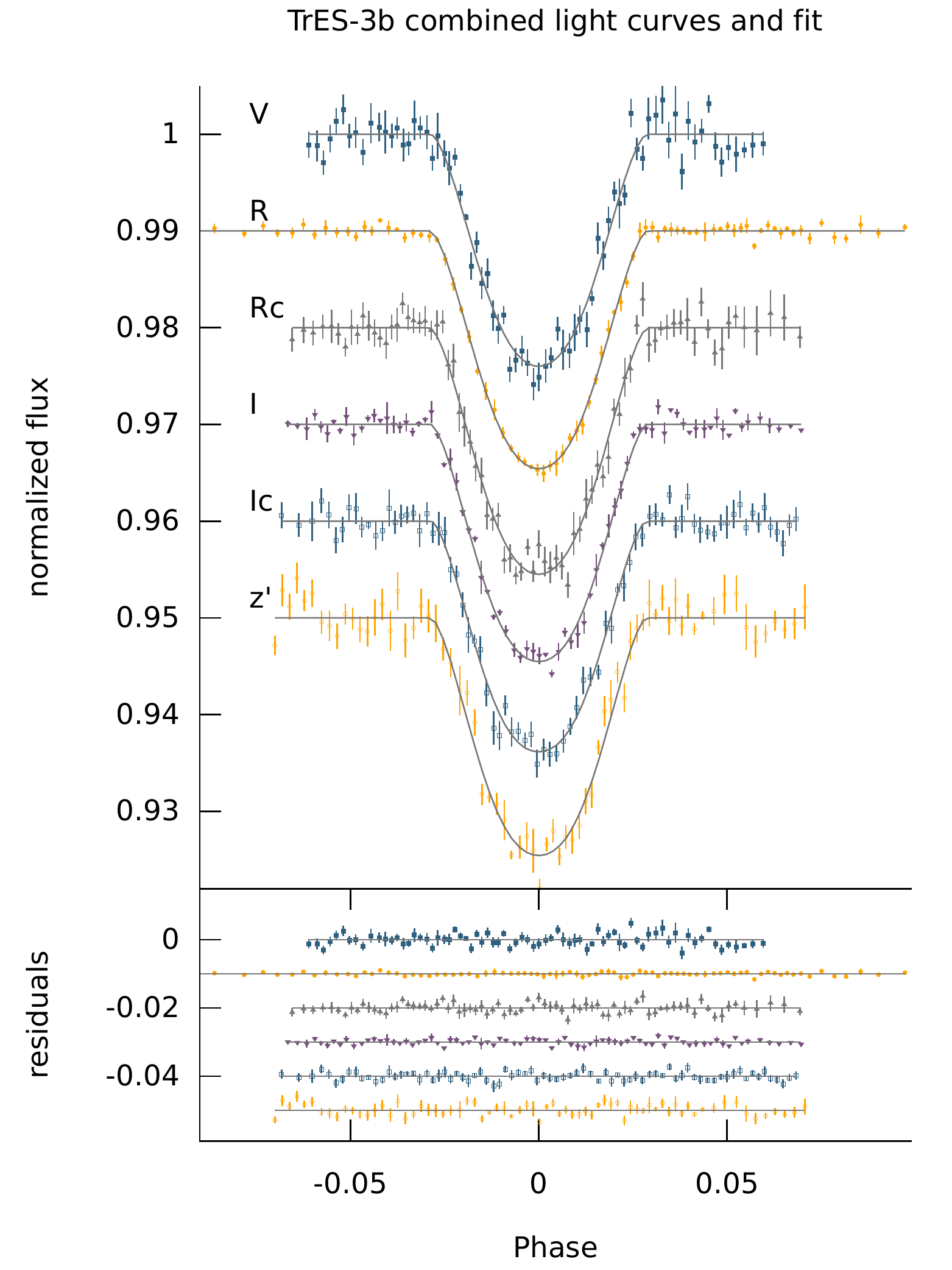}
  \caption{Light curves of \tr and fit.
    The lower panel shows residuals.
    \seeonline
    \label{tres-sup}}
\end{figure}

We used the same technique applied for \hh fitting for what concerns \tr
limb darkening, with following parameters: $log(g_*)=4.571\pm 0.006$
\citep{2011MNRAS.417.2166S}, $T_*=5650\pm75\kelvin$, and
$[Fe/H]_*=-0.19\pm0.08$ \citep{2009ApJ...691.1145S}.

\textsc{exofast} fit results of \tr are shown in Table.~\ref{fittr}.

\begin{figure}[t]
  \centering
  \includegraphics[width=\columnwidth]{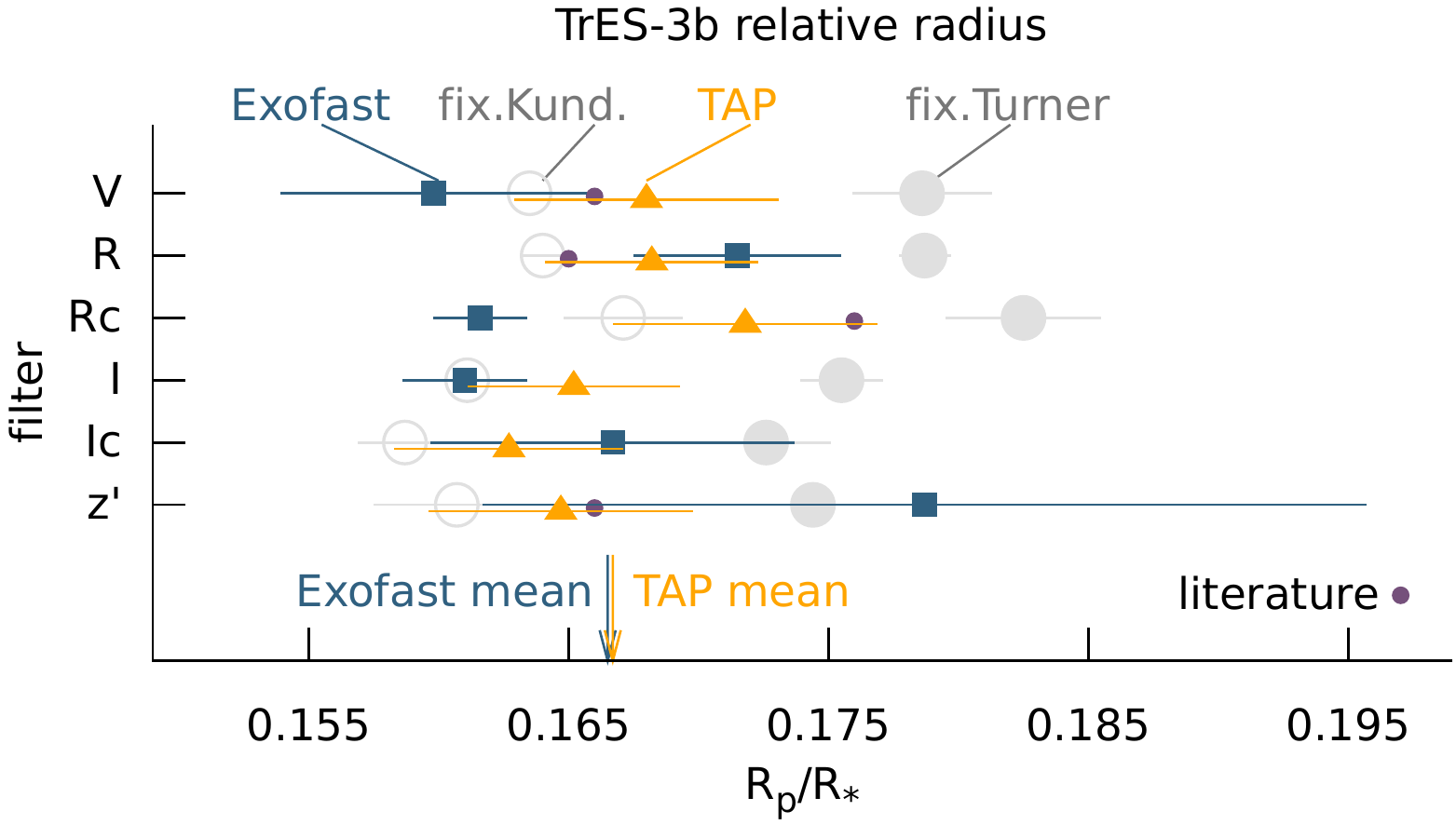}
  \caption{ $R_p/R_*$ values of \tr for each observing filter 
    resulting from \textsc{exofast} and \textsc{tap} fit procedures. 
    Arrows show weighted means.
    Large light circles show \textsc{tap} results
    obtained by fixing $i$ and $a/R_*$ with values provided by
    \cite{2013ApJ...764....8K} and \citep{2013MNRAS.428..678T}.
    Small bold circles show values provided in
    the literature:
    %
    $V$ and $Rc$ \citep{2013MNRAS.428..678T},
    $R$          \citep{2013ApJ...764....8K}, and
    $z'$         \citep{2007ApJ...663L..37O}.
    \seeonline
    \label{tres-radius}}
\end{figure}

While performing the \textsc{tap} fit, we fitted a unique value of $i$
and $a/R_*$ for all combined curves, and we fitted separately a value of
$R_p/R_*$ for each considered filter.  \textsc{tap} results show good
agreement with the \textsc{exofast} weighted averages of $i$ and
$a/R_*$:
$i = 81.70^{+0.17}_{-0.22}$ of \textsc{tap} against
$    81.63^{+0.20}_{-0.23}$ of \textsc{exofast}, and
$a/R_* = 5.885^{+0.068}_{-0.080}$ of \textsc{tap} against
$        5.903\pm{0.062}$ of \textsc{exofast}.
Results are slightly lower than, but in agreement with, the values
provided by \cite{2013ApJ...764....8K}:
$i     = 81.86^{+0.08}_{-0.26}$ and
$a/R_* = 5.91^{+0.04}_{-0.05}$.
Concerning $R_p/R_*$, we have differences between the two fit
procedures which are within $-2\sigma$ (for the $V$ filter) and
$+1\sigma$ (for the $I$ filter).
Comparisons of $R_p/R_*$ between different filters show a maximum
absolute variation of $2.9\sigma$ between $R$ and $V$ curves in
\textsc{exofast} results, which are not confirmed by \textsc{tap}
results ($0.04\sigma$). We then assume no $R_p/R_*$ variations with
the observing filters, and we consider average results finding good
agreement between fit procedures:
$R_p/R_* = 0.1665^{+0.0081}_{-0.0050}$ for \textsc{exofast} and
         $ 0.1667^{+0.0047}_{-0.0035}$ for \textsc{tap}, against the value of
         $ 0.1649 \pm 0.0015$ provided by 
\cite{2013ApJ...764....8K} in the $r'$ band (see Fig.~\ref{tres-radius}).

%
Fig.~\ref{tres-radius} also show additional \textsc{tap} fit solutions
obtained by fixing $i$ and $a/R_*$ found by
\cite{2013ApJ...764....8K} and \citep{2013MNRAS.428..678T},
and fitting only a separate value of $R_p/R_*$ for each
observing filter.
%

\begin{figure}[b]
  \centering
  \includegraphics[width=\columnwidth]{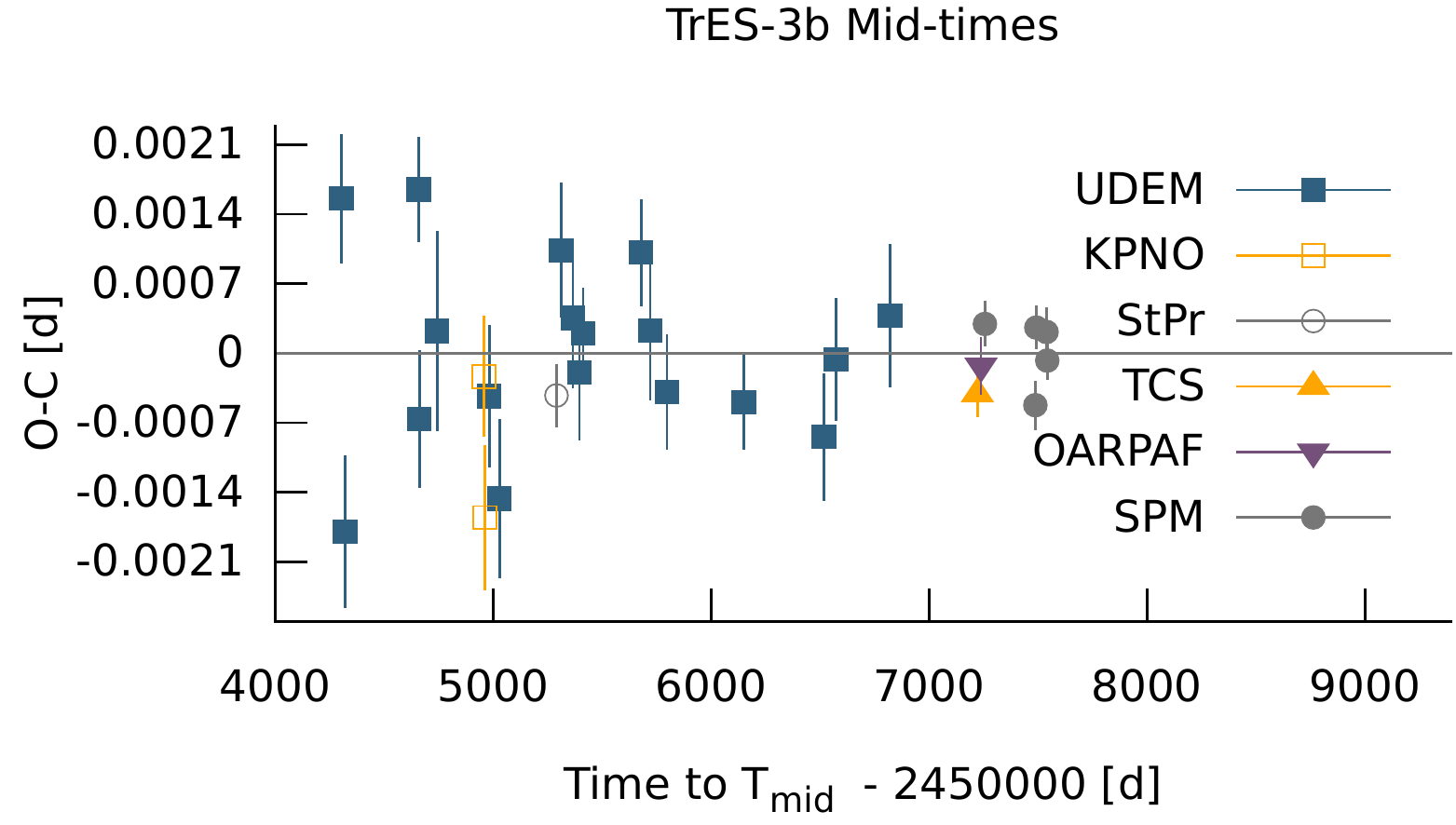}
  \caption{Observed minus Calculated Mid-times values for \tr
    observations presented in this work. Data are consistent
    with a $P=1.3061862\pm0.0000001$.
    \seeonline
    \label{tres-midtimes}}
\end{figure}

Mid-times fit linearly a period of $P = 1.3061862\pm0.0000001$ days
(see Fig.~\ref{tres-midtimes}). Scatter of calculated values of
$T_{mid}$ corresponds to variations of $\approx 4$ minutes
peak-to-valley, mostly ruled by UDEM data, which is reduced to
$\approx 1$ minute peak-to-valley while considering OAN-SPM, OARPAF
and TCS data. Calculated Mid-times show no significant linear
variation of the period with the time, and our sampling does not allow
us to search for periodic variations in the $1:2$ resonance as in the
work of \cite{2013ASInC...9...78T}.

\section{Conclusions}
\label{conc}

\begin{table}[t]
\centering
\caption{\textsc{exofast} fit results for \hh. 
  \label{fithh}}
\begin{tabular}{CcCCC}
\tableline                                      
\tableline                                      
  &  obs.   &           i           &         a/R_*         & R_p/R_*          \\
\tableline                                                                                       
B & KPNO-VCT& 86.78^{+0.29}_{-0.27} & 10.17^{+0.37}_{-0.34} & 0.1112^{+0.0025}_{-0.0026} \\
V & OAN-SPM & 86.57^{+0.27}_{-0.26} & 10.19^{+0.34}_{-0.32} & 0.1111^{+0.0018}_{-0.0018} \\
R & OAN-SPM & 86.83^{+0.36}_{-0.39} & 10.12^{+0.34}_{-0.36} & 0.1079^{+0.0014}_{-0.0012} \\
I & OAN-SPM & 86.71^{+0.29}_{-0.27} & 10.18^{+0.32}_{-0.31} & 0.1083^{+0.0015}_{-0.0015} \\
Ic&  UDEM   & 86.74^{+0.31}_{-0.29} & 10.18^{+0.32}_{-0.30} & 0.1094^{+0.0012}_{-0.0012} \\
z'& KPNO-VCT& 88.96^{+0.71}_{-0.90} & 10.04^{+0.16}_{-0.21} & 0.1129^{+0.0015}_{-0.0015} \\
JH&  KPNO   & 86.60^{+0.25}_{-0.24} & 10.16^{+0.29}_{-0.28} & 0.1084^{+0.0008}_{-0.0008} \\
\tableline 
\end{tabular}
\end{table}

\begin{table}[t]
  \centering
\caption{\textsc{exofast} fit results for \tr. 
  \label{fittr}}
\begin{tabular}{CcCCC}
\tableline                                      
\tableline                                      
   &  obs.   &           i           &          a/R_*          & R_p/R_*                   \\
\tableline                                      
V  &  UDEM   & 81.34^{+0.23}_{-0.25} & 5.903^{+0.066}_{-0.066} & 0.1598^{+0.0059}_{-0.0045} \\  
R  &   {a}   & 81.80^{+0.15}_{-0.17} & 5.899^{+0.054}_{-0.056} & 0.1715^{+0.0040}_{-0.0030} \\ 
Rc &  UDEM   & 81.59^{+0.15}_{-0.14} & 5.913^{+0.060}_{-0.059} & 0.1616^{+0.0018}_{-0.0019} \\  
I  & OAN-SPM & 81.77^{+0.14}_{-0.16} & 5.895^{+0.061}_{-0.061} & 0.1610^{+0.0024}_{-0.0015} \\ 
Ic &  UDEM   & 81.83^{+0.19}_{-0.22} & 5.902^{+0.063}_{-0.064} & 0.1667^{+0.0070}_{-0.0050} \\  
z' &   {b}   & 81.46^{+0.28}_{-0.37} & 5.906^{+0.067}_{-0.066} & 0.1787^{+0.0170}_{-0.0095} \\   
\tableline 
\end{tabular}
\tablenotetext{a}{Combining OAN-SPM, OARPAF, and TCS data.}
\tablenotetext{b}{Combining UDEM, KPNO, and StPr data.}
\end{table}

We obtained ten new exoplanetary transit observations of \hh in the
$BVRIz'JH$ bands, and twenty-six new observations of \tr in the
$VRIz'$ bands, which confirmed the potential adequacy of small-size
telescopes ($36$--$152\centi\meter$) for this research topic.  In
particular, the new OARPAF observatory and the new instrument \wf at
TCS can provide reliable photometric observations in the framework of
ground-based exoplanetary transit follow-ups.

The simultaneous fit of light curves, carried out in multiple
observing bands by several telescopes, allowed us to achieve orbital
and physical parameters which corroborate results of
\cite{2011AA...527A..85N} for what concerns \hh, and of
\cite{2013ApJ...764....8K} for what concerns \tr.

We also report specific values of $R_p/R_*$ in multiple optical and
near-infrared bands, and a first estimation of this parameter in the
$B$ band for \hh which is coherent with our values provided by using
other filters.

We find that observing filters do not significantly influence the
determination of the relative radius of \hh and \tr.  This result
confirms the relative radius of \hh estimated by near-infrared band
observations \citep{2009AA...493L..35D}, and the relative radius of
\tr estimated by UV observations \citep{2013MNRAS.428..678T} in the
literature.

Mid-times of \tr fit a $P = 1.3061862\pm 0.0000001$ days, finding no
significant linear variations of the period over $\approx 9$ years of
photometric observations.  Further observations are planned in order
to discriminate eventual $R_p/R_*$ variations with the observing
filter and to extend $T_{mid}$ information in order to estimate TTVs.

\acknowledgments

The authors acknowledge the following financial support:
DR from the Spanish Ministry of Economy and Competitiveness (MINECO)
under the 2011 Severo Ochoa Program MINECO SEV-2011-0187;
LFM and RM from the UNAM under grant PAPIIT IN 105115;
DR and FGRF from the CONACYT scholarship for postgraduate studies in
Mexico.
Research carried out thanks to the support of UNAM-DGAPA-PAPIIT
project IN115413.
SN thanks H.~Navarro-Meza for assistance during OAN-SPM observations
in 2016/06.
DR thanks Claudia Molina and Katie Marley for language editing, and
the anonymous referee for remarks.
We also acknowledge the OAN staff for daily support.
The $1.5\meter$ Carlos S\'anchez Telescope is operated on the island of
Tenerife by the Instituto de Astrof\'isica de Canarias in the Spanish
Observatorio del Teide.

{\it Facilities:} 
\facility{
  OAN-SPM $84\centi\meter$ (MEXMAN),
  UDEM $80\centi\meter$,
  TCS $152\centi\meter$ (\wf),
  OARPAF $80\centi\meter$
}

{\footnotesize
\bibliographystyle{aasjournal} 
\bibliography{biblio}{}
}


\end{document}